\documentstyle[11pt]{article}
\def \ea{{\it et al.}}
\def \ie{{\it i.e.}}
\def \eg{{\it e.g.}}

\let\a=\alpha
\let\b=\beta
\let\d=\delta

\let\k=\kappa
\let\l=\lambda
\let\L=\Lambda
\let\m=\mu
\let\n=\nu
\let\r=\rho

\let\s=\sigma

\let\w=\omega
\let\O=\Omega

\def \lb{\left[}              %
\def \rb{\right]}             

\def \ld{\left\langle}
\def \rd{\right\rangle}

\def \cC{{\cal C}}

\def \cG{{\cal G}}
\def \cH{{\cal H}}
\def \cL{{\cal L}}
\def \cN{{\cal N}}

\def \cP{{\cal P}}
\def \cQ{{\cal Q}}

\def \cZ{{\cal Z}}
\def \vec #1{\mbox{\boldmath ${#1}$}}
\def \svec #1{\mbox{\boldmath ${\scriptstyle #1}$}}
\def \be{\begin{equation}}
\def \ee{\end{equation}}
\def \bea{\begin{eqnarray}}
\def \eea{\end{eqnarray}}

\begin{document}
\title{Equilibrium Ensemble Approach to Disordered Systems I:
                      General Theory, Exact Results}
\author{Reimer K\"uhn\thanks{Supported by a Heisenberg fellowship}\\
Institut f\"ur Theoretische Physik, Universit\"at Heidelberg\\
Philosophenweg 19, 69120 Heidelberg, Germany\\
{\small e--mail: kuehn@hybrid.tphys.uni-heidelberg.de}\\ }
\date{May 1995}
\maketitle
{\bf Abstract}\  An outline of Morita's equilibrium ensemble approach to
disordered systems is given, and hitherto unnoticed relations to other,
more  conventional approaches in the theory of disordered systems are
pointed out. It is demonstrated to constitute a generalization of the
idea of grand ensembles and to be intimately related also to conventional
low--concentration expansions as well as to perturbation expansions about
ordered reference systems. Moreover, we draw attention to the variational
content of the equilibrium ensemble formulation. A number of exact results
are presented, among them general solutions for site-- and bond-- diluted
systems in one dimension, both for uncorrelated, and for correlated
disorder.

\bigskip

\section{Introduction}

The present contribution is concerned with the statistical mechanics of
systems with quenched randomnes. According to a seminal paper by Brout
\cite{brou},  the relevant thermodynamic potential for the description of
such systems is the average of the free energy of the system  over the
distribution $q(\k)$ describing the disorder configurations $\k$,
\be
f_q = -(\b N)^{-1} \ld \ln Z_N(\k)\rd_{q}\ .
\label{fqu}
\ee
Here $N$ denotes the system size, and $Z_N(\k)$ is the partition function
of a system of size $N$ at fixed disorder configuration $\k$.

The evaluation of the quenched free energy (\ref{fqu}) is difficult for at
least two reasons. First, in virtually all configurations $\k$ contributing
to  the average, the system is not translationally invariant ---
a circumstance invalidating all conventional calculational tools relying
on such invariance. Moreover, averaging the logarithm of the partition
function usually precludes any useful factorization of the averaging
process even in situations where such factorization is possible when
considering the average of $Z_N(\k)$ itself.

For the reasons just outlined only few exactly solved models of systems
with quenched disorder are known. Where solutions of such models {\it
have\/} been obtained, success has as a rule been due to simplifying
features, such as one--dimensionality of the models (see \cite{jml} for a
recent overwiew) or of the disorder type  \cite{mcwu}, analytic structures
available in spherical models \cite{past} or infinite dimensionality where
mean--field approximations are exact (see, e.g. \cite{mpv}).

Discomforting to a perhaps even greater degree is the fact that conventional
asymptotic methods for the study of phase transitions and critical phenomena,
such as series expansions, renormalization group or Monte--Carlo calculations,
have in several cases procuced conflicting and sometimes even spurious
results, when applied to disordered systems. This being so, alternative
approaches are clearly welcome.

The present contribution is devoted to one such alternative, which is
at the same time venerable, {\em not\/} well known and still, we believe,
promising. The method is originally due to Morita \cite{mor}. The main idea
is to treat the configurational degrees of freedom $\k$ on the same footing
as the dynamical variables proper, and to supplement the Hamiltonian
of the disordered system by a disorder potential $\phi(\k)$ which is
determined such that configuration averaging as implied by (\ref{fqu})
becomes part of the Gibbs average in an enlarged phase space. By this
device, it is hoped, the full machinery of equilibrium statistical physics
is made available for the study of systems with quenched randomness. The
idea of an equivalent equilibrium ensemble naturally lends itself to the
formulation of systematic approximative schemes. The perhaps most natural
was already suggested by Morita \cite{mor}; it starts out from the
so--called annealed approximation, which is then improved by imposing an
increasing number of constraints on the thermal motion of the
configurational degrees of freedom. Thus, as pointed out by Falk \cite{fa},
annealed approximations, such as employed by Thorpe and Beeman \cite{tb}
to determine phase boundaries in bond--disordered 2--D Ising models,
constitute just the first step wihtin Morita's restricted annealing scheme.

The Morita approach is natural enough to have been used, reconsidered, or
rediscovered several times. Thorpe \cite{tho} applied essentially this
idea to improve upon annealed phase boundaries obtained earlier \cite{tb},
a topic taken up again somewhat later by George et al.\/ \cite{geo+}, who
fell on virtually the same set of ideas, arguing from a somewhat different
point of view, though. Restricted annealing has been discussed in connection
with spin--glasses by Toulouse and Vannimenus \cite{touva}, who also drew
attention to formal similarities with lattice gauge theories.

Explicit reference to Morita's original paper is made in \cite{swa} --
\cite{kuprl}. Sobotta and Wagner \cite{swa}, restricting themselves to
bimodal disorder (see below), streamlined the formal presentation
of the approach, and applied it in conjunction with the renormalization
group and the $\varepsilon$--expansion to study spin--diluted ferromagnets
near
dimension 4 \cite{swab}. Huber \cite{hu} discusses the {\it thermodynamic\/}
content of the Morita scheme and derives a normalization condition for the
disorder potential $\phi(\k)$ from thermodynamic consistency--considerations.
In \cite{kdiss}, \cite{kuprl}, K\"uhn combines restricted annealing with
the phenomenological renormalization group method to study the critical
behaviour of the 2--D spin--diluted Ising model, while K\"uhn \ea\/
\cite{ku+} apply constrained annealing to provide an exact solution of
van Hemmen's spin--glass model \cite{vh}.

Somewhat earlier restricted annealing was rediscovered by Schwartz
\cite{sch} and put forward in the debate about the lower critical dimension
of the random field Ising model (RFIM), incidentally producing results
compatible with the more recent rigurous proof of stability of ferromagnetism
in the three dimensional RFIM \cite{bric+}. The most recent rediscovery of
the method is due to Serva and Paladin \cite{sepa}, who used it to study
the mean--field version of the RFIM, which is actually somewhat simpler
that the model considered in \cite{ku+}, and who produced approximate
numerical solutions of the RFIM in 1--D.

The purpose of the present  paper is to give an outline of the formal and
systematic aspects of Morita's equilibrium ensemble approach to disordered
systems. Sec. 2 summarizes for future reference the general theory, and
formulates constrained annealing as the canonical approximative scheme that
derives from it. In this context it is emphasized that the equilibrium
ensemble approach is nothing but a generalization of the grand ensemble
idea \cite{kuprl}. Alternatively, it can be looked upon as a method for
obtaining variational lower bounds for quenched free energies.
In Sec. 3 we specialize to bimodal bond-- or site--disorder, and point out
a connection to a different well established set of ideas, namely the
relation to conventional low--concentration expansions and to perturbation
expansions about homogeneous reference systems. For systems with bond-- or
site--dilution additional information about the exact disorder potential
$\phi(\k)$ can be obtained which is used among other things to provide
exact general solutions to the 1--D case. The equilibrium ensemble
approach is well suited to study systems with correlated
quenched disorder. Sec.~4 develops this point to some extent. In Sec.~5
we briefly indicate a range of alternative possibilities to formulate
truncation and approximation schemes within the general approach, and
we end with an outlook and concluding remarks. Constrained annealing as a
method to study the thermodynamics and critical behaviour of disordered
Ising models will be taken up in greater detail in an accompanying paper
\cite{II}, to be referred to as II in what follows. A short version
collecting the main results of that study has appeared elsewhere
\cite{kuprl}. The contents of Sec.~2 are not entirely new. Large parts of
the general theory, in particular Secs.~2.1--2.4, have more or less
explicitly appeared before --- notably in \cite{mor}, \cite{swa}.
What we can perhaps claim originality for, is for pointing out heretofore
unnoticed connections to other sets of ideas (to grand canonical ensembles,
to variational bounding of free energies, and to low--concentration and
perturbation expansions) as well as for the collection of exact results
presented in Secs. 3 and 4.

\section{General Theory}

\subsection{Equilibrium Ensemble Approach to Disordered Systems}

In order to avoid the complications associated with the evaluation of
quenched averages, Morita \cite{mor} suggested to introduce an equivalent
ensemle
in which configurational averaging becomes part of the thermal averaging
procedure in an enlarged phase space that is defined over both, the
configurational degrees of freedom $\k$, and the dynamical variables
proper. This is achieved by introducing a ``ficticious" disorder
potential $\phi(\k)$
which is added to the system Hamiltonian
\be
\cH(\s|\k) \to \cH^\phi(\s,\k) = \cH(\s|\k) + \phi(\k)
\label{fhamilton}
\ee
and chosen such that the system described by $\cH^\phi$ will have
thermodynamic equilibrium properties identical to the nonequilibrium
properties of the quenched system.\footnote{Here and in what follows, we
suppress indices signifying the system--size dependences of various
quantities, in order to simplify notation.} In a first step, therefore, it
is required that the Gibbs-Boltzmann distribution generated by $\cH^\phi$
\be
p^\phi(\s,\k) = \frac{1}{Z^\phi} \exp[-\b \cH^\phi(\s,\k)]
\label{fgibbs}
\ee
be equal to the non--equilibrium distribution
\be
p_q(\s,\k) = \frac{q(\k)}{Z(\k)} \exp[-\b \cH(\s|\k)]
\label{pqbrout}
\ee
describing the joint statistics of configurational and dynamical
degrees of freedom in an ensemble of quenched systems. That is,
one demands
\be
p^\phi(\s,\k) = p_q(\s,\k)
\ee
for all $(\s,\k)$. This equation can be solved for $\phi$ to yield
\be
\b \phi(\k) = - \ln [q(\k)/Z(\k)] - \ln Z^\phi\ .
\label{solvephi}
\ee
We shall in what follows refer to (\ref{solvephi}) as to the Morita
equation.
It expresses the fact that an equivalent equilibrium ensemble exists.
As long as one is interested only in its probabilistic content,
Eqs.~(\ref{fhamilton}), (\ref{fgibbs}) and (\ref{solvephi}) are all one
needs in principle. In practice, of course, not much has been gained so
far. The translation of the problem of configuration averaging in systems
with quenched randomness into the language of an equivalent equilibrium
ensemble is {\it purely formal}. It does not so far help to circumvent any
of the difficulties for which it was invented to begin with. To use the
idea in practice, one has to find a {\it representation\/} for the
potential $\phi(\k)$, which is adapted to the given disorder problem,
and which can be the starting point for the construction of systematic
approximation schemes.

Before turning to this topic in the following subsection, let us address
the thermodynamic content of the equivalent equilibrium ensemble.
It was pointed out by Huber  \cite{hu}, that due to the invariance of
(\ref{solvephi}) under the transformation
\bea
\phi(\k) & \to & \phi(\k) + \phi_0\ , \\
Z^\phi   & \to & Z^\phi \exp[- \b\phi_0]\ ,
\eea
the disorder potential $\phi$ and thus the equilibrium ensemle's partition
function
$Z ^\phi$ are specified only up to arbitrary constants $\phi_0$ (constant
in the sense that they do not depend on the disorder configuration $\k$) ,
which may, however, still depend on temperature, external fields, coupling
constants, impurity concentration and so on. For the probabilistic content
of the  theory, this need not be a point of concern. If, however, one also
demands that
\be
F^\phi = -\b^{-1} \log Z^\phi
\ee
be identified with the free energy of the equivalent equlibrium ensemble,
a normalization is clearly needed. It was demonstrated by Huber \cite{hu}
that thermodynamic consistency --- equality of probabilistic and
thermodynamic definitions of internal energy, entropy, and other
thermodynamic functions involving derivatives of $F^\phi$ up to first order
in temperature and fields --- requires the configuration average of $\phi$
to vanish,
\be
\ld \phi(\k) \rd_q = \sum_\k q(\k) \phi(\k) = 0\ .
\label{zmpc}
\ee
Indeed, with this condition it follows from (\ref{solvephi}) that
\be
-\b F^\phi = \ln Z^\phi = \ld \ln Z(\k) \rd_q - \ld \ln q(\k) \rd_q\ ,
\label{ffree_en}
\ee
so that $F^\phi$ gives the quenched free energy plus a contribution which
is easily identified with an entropy of mixing,
\be
k_{\rm B}^{-1} S_0 = -\ld \ln q(\k) \rd_q = - \sum_\k q(\k) \ln q(\k) \ .
\ee
Since this extra contribution is independent of temperature and external
fields, it is irrelevant for thermodynamics. With the normalization
condition imposed, the disorder potential $\phi$ reads
\be
\b \phi(\k) = -\lb \ln \frac{q(\k)}{Z(\k)} - \ld \ln \frac{q(\k)}{Z(\k)}
\rd_q \rb\ ,
\label{fullphi}
\ee
which shows that a full specification of $\phi$
requires the computation of the quenched free energy  --- precicely the
quantity the computation of which the equilibrium ensemble approach was
invented to facilitate, not to speak of the difficulties of solving or
analysing models with a contribution to the potential energy as complicated
as (\ref{fullphi}). On the other hand Eq. (\ref{fullphi}) shows that,
having computed a properly normalized $\phi(\k)$, it suffices to evaluate
it for a simple homogeneous configuration $\k$ to obtain essentially the
Brout free energy.

\subsection{Constrained Entropy Maximization}

In order to introduce an alternative and more explicit representation of
the disorder potential $\phi$, which can serve as a starting point for
systematic approximation schemes, it is advantageous to recall Mazo's
\cite{maz}
information-theoretic justification of Brout's averaging prescripton.

The distribution $p_q(\s,\k)$ which reproduces Brout's averaging
prescription for the free energy can according to Mazo be derived from
Jaynes' information--theoretic approach to statistical mechanics
\cite{jay}.
That is, it is given as the unique distribution which maximizes the
entropy functional
\be
 S[p] = -k_{\rm B} \sum_{\s,\k} p(\s,\k) \ln p(\s,\k)
 \label{entrop}
\ee
under the constraints
\bea
\ld \cH(\s|\k) \rd_p & = \sum_{\s,\k} p(\s,\k) \cH(\s|\k) & =  E
\label{constrE}\\
\ld \d(\k - \k')\rd_p & = \sum_{\s,\k} p(\s,\k) \d(\k - \k') & =  q(\k')\
,\qquad \forall\ \k' \ . \label{constrq}
\eea
Eq.~(\ref{constrE}) fixes the average energy of the system, while
(\ref{constrq}) demands that the search for maximizing distributions
is restricted among those compatible with the a--priori distribution
$q(\k)$ characterizing the disorder. Introducing Lagrangian multipliers
$-k_{\rm B}\b$ and $k_{\rm B}(\l(\k)+1)$ to enforce the constraints, one
obtains $p_q(\s,\k)$ defined by
\be
S[p_q] = \max_p \Big\{S[p] ;\ \ld \cH(\s|\k) \rd_p=E\ {\rm and}\ \ld \d(\k
- \k')\rd_p = q(\k')\ \forall\ \k'
\Big\}
\ee
in the form
\be
p_q(\s,\k) = \exp[-\b \cH(\s,\k) + \l(\k)]\ .
\label{pqmazo}
\ee
To satisfy (\ref{constrq}), one on needs
\be
\exp[\l(\k)] = \frac{q(\k)}{Z(\k)}\ ,
\ee
where $Z(\k)$ is the partition sum at fixed disorder configuration $\k$,
so that (\ref{pqmazo}) does, indeed, agree with (\ref{pqbrout}), and we
have seen in the previous subsection that it can be interpreted as an
equilibrium distribution $p^\phi$ in an enlarged phase space.

\subsection{Alternative Formulation of Constrained Entropy Maximization
(Bimodal Disorder)}

The alternative representation of $\phi$ announced above is obtained by
replacing the constraints (\ref{constrq}) in the above entropy--maximization
procedure by an equivalent set of constraints, namely by the requirement
that the search for maximizing distributions is restricted to those which
reproduce the {\it complete set of moments\/} of $q(\k)$.

To be specific, and to keep notation and formulae as trasparent as
possible, let us first consider the case  of systems with {\it bimodal\/}
site (bond) disorder, in which a random variable associated with each site
(bond) of the lattice can only take two values. We shall explain the
necessary modifications to treat more general types of randomnes in Sec.
2.5 below. Bimodal disorder is
conveniently described in terms of site  (bond) {\it occupation numbers\/}
$k_a=k_a(\k)$ which take the values 0 or 1, if on the $a$--th site (bond)
one or the other of the two possible realizations of the disorder is
attained in the configuration $\k$. The set of moments of $q(\k)$ is
given by the expectation under $q(\k)$ of all possible products of occupation
numbers $k_a$,
\be
f_\w = \ld \prod_{a\in \w} k_a(\k) \rd_q\ ,\qquad \w \subseteq \L\ ,
\label{moments}
\ee
where $\w$ ranges through all subsets of the set $\L$ of sites (bonds) of
the lattice, including the empty set. It is useful introduce occupation
numbers for the sets $\w$ through
\be
k_\w (\k) = \prod_{a\in \w} k_a(\k)\ ,
\ee
with $k_\emptyset(\k) \equiv 1$, and to identify $\k$ with the set of
vertices (bonds) for which $k_a=1$, that is, $\k = \{k_a; a\in\L\}\equiv
\{a\in\L; k_a=1\}$. This allows to write
\be
f_\w = \ld k_\w(\k) \rd_q = \sum_{{\k \atop \k \supseteq \w}} q(\k)\ .
\label{fq}
\ee
In the present case, the equivalence of the system of moments  $f_\w$ and
the distribution $q(\k)$  follows directly from the invertibility of
(\ref{fq}),
\be
q(\k) = \sum_{{\w \atop \w\supseteq \k}} (-1)^{|\w|-|\k|} f_\w \ ,
\label{qf}
\ee
where $|\w|$ and $|\k|$ denote the size of the sets $\w$ and $\k$,
respectively. The set of moments can be used to reformulate the constrained
entropy maximization, by replacing (\ref{constrq}) with
\be
\ld k_\w(\k) \rd_q = \sum_{\s,\k} p(\s,\k) k_\w(\k) = f_\w\ ,\qquad
\w\subseteq \L\ . \label{constrf}
\ee
Since (\ref{constrq}) and (\ref{constrf}) are {\it equivalent\/} sets of
constraints, one can alternatively obtain $p_q(\s,\k)$ through
\be
S[p_q] = \max_p \Big\{ S[p] ;\ \ld \cH(\s|\k) \rd_p=E\ {\rm and}\  \ld
k_\w(\k) \rd_p = f_\w\ ,\ \w\subseteq \L
\Big\}\ .
\ee
Denoting the Lagrangian multipliers that enforce (\ref{constrE}) and
(\ref{constrf}) by $-k_{\rm B} \b$  and $k_{\rm B} (\l_\w +
\d_{\w,\emptyset})$, one obtains the maximizing distribution $p_q$ in the
form \be
p_q(\s,\k) =  \exp\Big[ -\b \cH(\s|\k) + \sum_{\w\subseteq \L}
\l_\w k_\w(\k)\Big]\ .
\ee
Writing
\be
\l_\emptyset  =  - \ln \cZ ,
\label{lbdaempty}
\ee
one gets
\be
p_q(\s,\k) = \frac{1}{\cZ}\exp\Big[-\b \cH(\s|\k) + \sum_{\emptyset
\ne\w\subseteq\L} \l_\w k_\w(\k)\Big]\ ,
\label{altpq}
\ee
where the $\l_\w$, $\w \ne \emptyset$, must be determined such
that $p_q$ satisfies (\ref{constrf}) for the set of moments of the
distribution $q(\k)$ describing the disorder. Thus, an explicit
representation of $\phi(\k)$ in terms of weighted sums of products
of occupation numbers has been obtained. The representation (\ref{altpq})
of the distribution $p_q(\s,\k)$ has been advanced already by Morita
\cite{mor}, in a more transparent way, though, by Sobotta and Wagner
\cite{swa}.

Note that  $\cZ$ has the formal appearance of a generalized grand
partition sum, with generalized chemical potentials given by
$\mu_\w = \b^{-1} \l_\w$. The corresponding generalized grand
potential is
\be
\O = -\b^{-1} \ln \cZ\ ,
\ee
and one has
\be
f_\w = \ld k_\w \rd = - \frac{\partial \O}{\partial \mu_\w}\ .
\ee
Returning to the general considerations of the previous subsection,
we realize that in the form in which the disorder potential appears in
(\ref{altpq}),
\be
\b \phi(\k) = - \sum_{\emptyset \ne\w\subseteq\L} \l_\w k_\w(\k)\ ,
\ee
it does {\it not\/} satisfy the normalization condition (\ref{zmpc}).
To enforce it, one changes $\phi$ according to
\be
\b\phi(\k) \rightarrow  - \sum_{\emptyset \ne\w\subseteq\L} \l_\w
\big( k_\w(\k) - f_\w \big)\ .
\label{phizmpc}
\ee
On the level of thermodynamic potentials, this change amounts to
a Legendre--transform from the generalized grand potential to the
corresponding free  energy,
\be
\O \rightarrow F^\phi = \O + \sum_{\emptyset \ne\w\subseteq\L} \m_\w f_\w
= \O - \sum_{\emptyset \ne\w\subseteq\L} \m_\w \frac{\partial \O}
{\partial \mu_\w}\ .
\label{lagrtrans}
\ee

Finally, using (\ref{lbdaempty}), (\ref{lagrtrans}) and the fact that
$f_\emptyset =1$ due to the normalization of $p^\phi=p_q$, one obtains
an intriguingly simple representation of the equivalent ensemble's free
energy, viz.
\be
-\b F^\phi = \ln Z^\phi = - \sum_{\w\subseteq\L} \l_\w  f_\w\  .
\label{falt}
\ee
Note that determining the $\l_\w$ such as to make $p^\phi$ satisfy the
constraints defining the disorder distribution is tantamount
to {\em minimizing\/} $\ln Z^\phi$ with respect to the $\l_\w$. Phrased
differently, it means minimizing $\ln Z^\phi$ over a space of functions
$\phi(\k)$ which are of the form (\ref{phizmpc}).

\subsection{A Canonical Approximative Scheme: Constrained Annealing
(Bimodal Disorder)}

The simplicity of (\ref{falt}) is, of course, purely formal and hence
deceptive. Nevertheless, something non--trivial has been gained, namely
an {\it explicit\/} representation for the disorder potential $\phi$,
together with a method to compute it that immediately lends itself to
the formulation of a systematic scheme of approximations. This scheme
consists of replacing the full set (\ref{constrf}) of constraints
by a subset which requires the maximizing distributions to reproduce
only a given subset of the full set of moments of $q(\k)$.

Formally, let $\cQ$ be a subset of the powerset $\cP(\L)$ of the vertices
or bonds of the lattice, and define $p^\cQ$ as the unique distribution
that maximizes the entropy functional (\ref{entrop}) under the constraints
\bea
\ld \cH(\s|\k) \rd_p & = \sum_{\s,\k} p(\s,\k) \cH(\s|\k)  & = E
\label{constrEE}\\
\ld k_\w(\k) \rd_p & = \sum_{\s,\k} p(\s,\k) k_\w(\k) \quad  & = f_\w\
,\quad \w \in \cQ\ , \label{constrfQ}
\eea
that is,
\be
S[p^\cQ] = \max_p \Big\{ S[p] ;\ \ld \cH(\s|\k) \rd_p=E\ {\rm and}\ \ld
k_\w(\k) \rd_p = f_\w\ , \ \w\in\cQ
\Big\}\ .
\ee
Introducing Lagrangian multipliers $-k_{\rm B} \b$  and $k_{\rm B}
(\l_\w^\cQ + \d_{\w,\emptyset})$, $\w \in \cQ$, one obtains the maximizing
distribution $p^\cQ$ in the form
\be
p^\cQ(\s,\k) =  \exp\Big[ -\b \cH(\s|\k) + \sum_{\w\in \cQ}
 \l_\w^\cQ k_\w(\k)\Big]\ .
\ee
Writing
\be
\l_\emptyset^\cQ = - \ln  \cZ^\cQ\ ,
\ee
one observes that this takes the form of a generalized grand--canonical
distribution \be
p^\cQ(\s,\k) = \frac{1}{\cZ^\cQ} \exp\Big[-\b \cH(\s|\k) + \sum_{\emptyset
\ne\w\in \cQ} \l_\w^\cQ k_\w(\k)\Big]\ ,
\label{pcQ}
\ee
where one has to determine the $\l_\w^\cQ=\b\mu^\cQ_\w$ such as to
satisfy the constraints (\ref{constrfQ}). In order to obtain the free
energy corresponding to the $\cQ$--approximation, one has to enforce the
zero--mean-potential condition (\ref{zmpc}) by changing $\phi^\cQ$
according to
\be
\b \phi^\cQ(\k) \rightarrow - \sum_{\emptyset\ne\w\in\cQ} \l_\w^\cQ
\big ( k_\w(\k \big) - f_w )\ .
\label{phizmpcQ}
\ee
As above, this amounts to a Legendre transformation from the generalized
grand potential
\be
\O^\cQ = -\b^{-1} \ln \cZ^\cQ
\ee
to the corresponding free energy
\be
\O^\cQ \rightarrow F^\cQ = \O^\cQ + \sum_{\emptyset \ne\w\in\cQ} \m^\cQ_\w
f_\w = \O^\cQ - \sum_{\emptyset \ne\w\in\cQ} \m^\cQ_\w \frac{\partial \O^\cQ}
{\partial \mu^\cQ_\w}\ .
\label{FcQ}
\ee
Once more, we have a simple representation for the free energy,
\be
-\b F^\cQ = \ln Z^\cQ = - \sum_{\w\in\cQ} \l_\w^\cQ  f_\w\  .
\label{fQ}
\ee
Again, to determine the $\l_\w^\cQ$ so as to make $p^\cQ$ satisfy the
restricted set (\ref{constrfQ}) of constraints is equivalent to minimizing
$\ln Z^\cQ$ with respect to these parameters or, put differently, over a
space of functions $\phi(\k)$ which are of the form (\ref{phizmpcQ}).

Unlike before, the simplicity of (\ref{fQ}) need not be purely formal.
For instance, if one restricts one's attention to simple approximations
which reproduces only a small subset of the set of moments of $q(\k)$
---  say the translationally invariant expectations $\ld k_a\rd$ and
the translationally invariant nearest neighbour  correlations $\ld
k_a k_b\rd$ in the case of bimodal site--disorder, so $\cQ=\{\{a\}_{a\in\L},
\{(a,b)\}_{a,b\in\L}\}$, with $(a,b)$ denoting nearest neighbours --- then
the corresponding Hamiltonian $\cH^\phi$ acquires only on-site terms and
nearest neighbour interactions between the $k$. So one obtains a fairly
simple translationally invariant model. For its study the full machinery
of equilibrium statistical mechanics and an established arsenal of
approximation methods is, at least in principle, available. We shall
demonstrate below that such simple approximations can already yield
very precise, if not exact, results.

There are a number of requirements the set $\cQ$ should meet to give rise
to a sensible approximation for the underlying problem of describing
systems with quenched disorder. First, to ensure that the distribution
$p^\cQ$ is normalized, $\cQ$ must contain the empty set, since the
normalization constraint is expressed precisely by the condition $f_\emptyset
= \ld k_\emptyset\rd = 1$. Moreover, if the disorder is homogeneous in the
sense that the distribution $q(\k)$ has a set of moments invariant under
the group $\cG$ of operations which map the underlying lattice, hence
the set $\L$ of sites (bonds), onto itself, the set $\cQ$ should be
such that it respects these symmetries, i.e.,  it ought to be closed
under the operations $g$ of the symmetry group $\cG$ of the underlying
lattice. Formally, if $\w\in \cQ$  then $g\w = \{a\in\L;\ g^{-1}a \in
\w\}$ should also be a member of $\cQ$. The invariance of the set of
moments under $\cG$ entails a corresponding invariance of the Lagrangian
multipliers, $\l^\cQ_{g\w} = \l^\cQ_\w$ for all $g$ in $\cG$. In other
words, the set of constraints can be decomposed into equivalence classes
under $\cG$.

Fixing $\cQ$ so as to concentrate on moments of the disorder
distribution which are of finite order or which involve occupation
numbers referring to sites (bonds) only up to some maximal distance
obviously entails a notion of {\em locality\/} for the corresponding
disorder potential $\phi^\cQ(\k)$. In what follows, we shall therefore
refer to approximations of this type as $\cQ$--local.

\subsection{Exact Lower Bounds for the Quenched Free Energy}

Before turning to a more detailed evaluation of these ideas, let
us not fail to mention that the approximative scheme described above
provides {\it exact lower bounds\/} for the quenched free energy.

This is easily seen by realizing that the constrained entropy maximization
scheme described above can be substituted by a constrained minimization scheme
for the free energy functional
\be
F[p] = \sum_{\s,\k} p(\s,\k) \cH(\s,\k) + \b^{-1} \sum_{\s,\k} p(\s,\k)
\ln p(\s,\k)
\ee
using the set of moments $\ld k_\w \rd = f_\w$, $\w \in \cQ$ as constraints.
This yields
\be
F[p^\cQ] = \min_p \Big\{ F[p];\ \ld k_\w(\k) \rd_p = f_\w\ \ \w\in\cQ
\Big\}\ .
\ee
With suitable Lagrangian multipliers, the $p^\cQ$ so defined
coincides with (\ref{pcQ}) as obtained from the entropy maximization
scheme. If, moreover, one enforces the zero--mean--potential condition
(\ref{zmpc}), then $F[p^\cQ] = F^\cQ$, with $F^\cQ$ given by (\ref{fQ}).

The dependece of $F^\cQ$ on the set $\cQ\subseteq \cP(\L)$ is {\it
monotone\/} in the sense that
\be
F^\cQ \le F^{\cQ'} \le F^\phi \quad {\rm if}\ \cQ\subseteq \cQ' \subseteq
\cP(\L)
\ee
so that $F^\cQ$ approaches the quenced free energy $F^\phi$ monotonically
from below, as the set of constraints, as characterized by $\cQ$ is
increased. A different way to see this is to recall that the $\ln Z^\cQ$
are obtained by minimizing $\ln Z^\phi$  over a space of $\cQ$-local
functions $\phi$ --- a function space which becomes larger as the set
$\cQ$ is enlarged.

\subsection{Non--Bimodal and Continuous Disorder Distributions and the
Variational Content of the Equilibrium Ensemble Approach}

The modifications necessary to obtain explicit representations  of the
disorder potential and approximative schemes for non--bimodal or continuous
disorder distributions $q(\k)$ can be inferred from what has been outlined
in Secs.~2.3--2.4 above. We will briefly state them in what follows. They also
point --- more obviously than our previous considerations --- to the
variational content of the equilibrium ensemble approach to disordered
systems.

For the remainder of this section, we drop the restriction that the $k_a$
be bimodal. To begin with, let us assume for simplicity that {\it all\/}
moments of the (non--bimodal or continuous) disorder distribution $q(\k)$
exist. They are given as expectations under $q(\k)$ of all possible
products of powers of the $k_a$, \ie, by
\be
f_\w^{\svec n} = \ld \prod_{a\in\w} k_a^{n_a}(\k)\rd_q\qquad ,\ \w\in\L,\
n_a\in I\!\! N\ ,
\label{cmoments}
\ee
with ${\vec n} = (n_a)_{a\in\w}$. The partition function of the equivalent
equilibrium ensemble describing the disordered system is then given by
\be
Z^\phi = \sum_{\s,\k} \exp [-\b\cH(\s|\k) -\b \phi(\k) ]\  ,
\label{zzphi}
\ee
with $\phi$ of the form
\be
\b\phi(\k) = \sum_{\w\subseteq\L,{\vec n}} \l_\w^{\svec n} (k_\w^{\svec
n}(k) - f_\w^{\svec n})\ .
\label{phicont}
\ee
Here we have introduced
\be
k_\w^{\svec n}(\k) = \prod_{a\in\w} k_a^{n_a}(\k) \  .
\ee
In (\ref{zzphi}), (\ref{phicont}) we have chosen to satisfy the
zero--mean--potential condition (\ref{zmpc}) at the outset. The copling
constants $\l_\w^{\svec n}$ have to be determined such that the Gibbs
distribution $p^\phi$ corresponding to (\ref{zzphi}) reproduces the moments
of the underlying disorder distribution $q$, \ie, $\ld k_\w^{\svec n}\rd =
f_\w^{\svec n}$ is supposed to hold for all $\w$ and $\vec n$. This amounts
to minimize $\ln Z^\phi$ as a function of the $\l_\w^{\svec n}$, in
complete analogy with the bimodal case. As above, the $\l_\w^{\svec n}$ can
be interpreted as Lagrangian multipliers of a constrained entropy
maximiation scheme or, alternatively, of a constrained free energy
minimization scheme.

As before, approximations are obtained by enforcing only a subset of the
full set of constraints, that is, by requiring the Gibbs distribution
$p^\phi$ to reproduce only a subset of the full set of moments
(\ref{cmoments}) of the disorder distribution $q(\k)$. This can be done
in several ways, for instance by requiring a match with the $f_\w^{\svec
n}$
only for $\w$ in some subset $\cQ$ of the powerset $\cP(\L)$ of the set of
sites (or bonds) of the lattice, by matching moments only up to
some maximal value for the $n_a$, or by combining these two truncation
schemes in various ways.

If one defines the approximations through a moment matching scheme
restricted {\em only\/} with respect to the choice of the $\w$ in (\ref
{cmoments}) and (\ref{phicont}), demanding that $\w\in\cQ\subseteq\cP(\L)$
as in Sec.~2.4, the minimization of $\ln Z^\phi$ with respect to the
correspondingly restricted set of parameters $\l_\w^{\svec n}$, $\w\in\cQ$,
amounts to a minimization over a true infinite dimensional function
space, \ie, as a variational problem. It amounts to minimizing  $\ln Z^\phi$,
interpreted as a functional of $\phi$  over a space of functions $\phi$
which are $\cQ$--{\em local\/} in the sense defined above. That is, they
are of the form
\be
\phi(\k) = \sum_{\emptyset\ne\w\in\cQ} \Big( \psi_\w(\k) - \ld \psi_\w(\k)
\rd_q \Big) \ ,
\ee
each $\psi_\w$ being a function of the $k_a$, $a\in\w$, which is analytic
in every component, since power series expansions were assumed to exist;
cf. Eq. (\ref{phicont}). Restricting also the range of $n_a$ values would
further restrict the set of functions to multinomials of some maximal
degree.

Conversely, dropping the analyticity constraint on the $\psi_\w$ will
enlarge the function space sufficiently to treat cases where {\em not\/}
all moments of $q(\k)$ exist.

Having said this much, it should have become clear that the equivalent
equilibrium ensemble approach could have been formulated as a variational
problem right away. To wit, consider $Z^\phi$, interpreted as a functional
of $\phi$,
\be
Z^\phi = Z[\phi] =  \sum_\k  Z(\k) \exp[-\b\phi(\k)],
\ee
and let us assume at the outset that the $\phi$ under consideration satisfy
the zero--mean--potential condition (\ref{zmpc}); alternatively, one might
impose it a--posteriori.  Using the disorder distribution $q(\k)$ this may
be rewritten as an average:
\be
Z^\phi = Z[\phi] = \ld Z(\k) \exp[-\b\phi(\k) - \ln q(\k)] \rd_q\ .
\ee
By Jensen's inequality,
\be
\ln Z^\phi = \ln Z[\phi] \ge \ld \ln Z(\k)\rd_q  - \ld \ln q(\k)]\rd_q\ ,
\label{varia}
\ee
where we have invested $\ld \phi(\k)\rd_q =0$. The task then is to minimize
the left hand side of (\ref{varia}) over a suitable space of functions
$\phi$ satisfying $\ld \phi(\k)\rd_q =0$. This variational point of view
was utilized by George \ea\/ \cite{geo+} in their study devoted to the
computation of phase--boundaries of bond--disordered 2--D Ising models
---  without connecting it to the equilibrium ensemble idea and to the
heuristics associated with it, and without, so it seems, the awareness
that it allows to proceed to exact solutions.

Indeed, the inequality in (\ref{varia}) can be satisfied as an equality,
\ie, the lowest bound attained and thereby the exact quenched free energy
computed, if $\phi(\k)$ is chosen such that
\be
Z(\k) \exp[-\b\phi(\k) - \ln q(\k)] = {\rm  const}
\label{varmorita}
\ee
in the sense that this quantity does not depend on $\k$. The value of this
constant must, of course, be $Z^\phi$, so (\ref{varmorita}) is nothing but
the Morita equation (\ref{solvephi}) in disguise. That is, it is the formal
solution of the variational problem
\be
\frac{\d \ln Z[\phi]}{\d  \phi} = 0\ .
\ee

Approximations are obtained by restricting the space of functions $\phi$
over which a minimum of $\ln Z[\phi]$  is sought, in other words, the
domain of defenition of the functional $\ln Z[\phi]$, to functions which
are $\cQ$--local in the sense explained before.  Approximations can
systematically be improved by increasing the size of the set $\cQ$

\section{Recursive evaluation of coupling constants}

For systems with quenched, uncorrelated bimodal site or bond disorder,
one can write down more or less explicit expressions for the coupling
constants $\l_\w$ of the full disorder potential $\phi$. It turns out that
these expressions generate conventional low--concentration expansions
for the quenched free energy (\ref{fqu}) as well as perturbation expansions
about pure reference systems. In the present section, we shall establish
this connection. In Sec.~3.2, we specialize to models with bond-- or
site--dilution for which additional exact information about the
exact disorder potential $\phi(\k)$ is available. Among other things, we
use it to reproduce the known exact solutions of one--dimensional models
with quenched site or bond dilution and to exhibit simplifying features
of these solutions in the absence of symmetry breaking fields.

\subsection{General results}

We consider systems with quenched, uncorrelated bimodal site or bond
disorder, defined on a lattice $\cL$ containing $N$ sites. As before,
let $\L$ denote the set of sites (bonds) of the lattice. The distribution
$q(\k)$ describing the disorder configurations
\be
\k  = \{ k_a;\ a\in \L\} \equiv \{ a \in \L;\ k_a = 1\}
\ee
is characterized by a single parameter $\r$,
\be
q(\k) = \prod_{a\in \L} \r^{k_a}(1-\r)^{1-k_a} =  \rho^{|\k|} (1 -
\rho)^{|\L| - |\k|} \ ,
\ee
so that
\be
f_\w = \ld k_\w \rd_q = \rho^{|\w|}.
\label{rhomoments}
\ee
We use the Morita equation (\ref{solvephi}),
\be
\b \phi(\k) = - \ln [q(\k)/Z(\k)] - \ln Z^\phi
\label{morita}
\ee
and the representation
\be
\b \phi(\k) = - \sum_{\emptyset \ne\w\subseteq\L} \l_\w ( k_\w(\k) - f_w )
\label{phiexact}
\ee
for the exaxt potential $\phi(\k)$, to obtain the coupling constants
$\l_\w$. This is achieved by inserting (\ref{phiexact}) and (\ref{rhomoments})
into the Morita equation, and by considering this equation first for
$\k=\emptyset$, which gives
\be
\b \phi(\emptyset) = \sum_{\emptyset \ne\w\subseteq\L} \l_\w  \r^{|\w|}
= - |\L| \ln (1-\r) + \ln Z(\emptyset) - \ln Z^\phi\ ,
\label{morempty}
\ee
and then for $\k\ne\emptyset$, which yields
\bea
\b\phi(\k) & = & \b\phi(\emptyset) - \sum_{\emptyset \ne\w\subseteq\k}
\l_\w\nonumber \\
           & = & - |\L| \ln (1-\r) - |\k| \ln\lb \frac{\r}{1-\r}\rb +
                 \ln Z(\k) - \ln Z^\phi\ .
\label{mor_nonempty}
\eea
Subtracting (\ref{mor_nonempty}) from (\ref{morempty}), we get
\be
\sum_{\emptyset \ne\w\subseteq\k} \l_\w  = - |\k| \ln\lb \frac{1-\r}{\r} \rb
- \ln Z(\k) + \ln Z(\emptyset)\ ,\quad \forall\ \k\ne\emptyset\ .
\label{mornonempty}
\ee
To determine the $\l_\w$ individually, we introduce the quantities
$x_\w$, $\w\subseteq\L$, through
\be
Z(\k) = \prod_{\w\subseteq\k} x_\w
\label{defx}
\ee
or, equivalently
\be
\ln Z(\k) = \sum_{\w\subseteq\k} \ln x_\w\ ,
\ee
where $\k$ ranges through all subsets of $\L$ (including the empty set) and
the product in (\ref{defx}) is over all subsets of $\k$, including again
the empty set. Eq.~(\ref{defx}) may be interpreted as a {\it recursive
definition\/} of the $x_\w$; alternatively, one may invert this relation
directly to yield
\be
\ln x_\w = \sum_{\k\subseteq\w} (-)^{|\w|-|\k|}  \ln Z(\k)\ .
\label{solvex}
\ee
Inserting this into (\ref{mornonempty}), and recalling that (\ref{mornonempty})
is supposed to hold for {\it all\/} $\k\ne\emptyset$, we can immediately
conclude
\be
\l_\w = - \ln \lb  x_\w \Big(\frac{1-\r}{\r}\Big)^{\d_{|\w|,1}} \rb\ ,\quad
\w\ne\emptyset\ .
\label{exactcouplings}
\ee
This is the more or less explicit expression for the coupling constants
$\l_\w$ of the exact disorder potential $\phi$ announced above. Using this
result, and returning, once more, to the $\k=\emptyset$ version
(\ref{morempty})
of the Morita equation, we get an (again more or less explicit) expression for
the free energy of the equivalent equilibrium ensemble,
\be
\ln Z^\phi = -|\L|\big[\r\ln \r + (1-\r)\ln(1 - \r)\big]  +
\sum_{\w\subseteq\L} \r^{|\w|} \ln x_\w \ .
\label{fexpans}
\ee
The first contribution in (\ref{fexpans}) is readily identified as the
expected contribution of the entropy of mixing.  Using (\ref{solvex}),
the second contribution can be cast into the form
\be
\sum_{\w\subseteq\L} \r^{|\w|} \ln x_\w =
\sum_{\k\subseteq\L} \r^{|\k|}(1-\r)^{|\L| - |\k|} \ln Z(\k) \ .
\label{lowconc}
\ee
Up to the factor $-\b^{-1}$ this is just the Brout avarage (\ref{fqu})  of
the free energy for the case of quenched, uncorrelated bimodal site or bond
disorder as it should, since no approximations were involved in the above
considerations.

Clearly, we cannot expect to be able to evaluate (\ref{fexpans})
completely, if we are unable to do so in the more standard formulation
(\ref{ffree_en}). Nevertheless, we have gained something useful.

First, Eq.~(\ref{fexpans}) together with (\ref{solvex}) amounts to a formal
model--independent prescription for reorganizing the Brout average in such
a way that it is turned into an expansion ordered by increasing powers of
$\r$. That is, for systems with bimodal bond-- or site--disorder, there is
apparently an intimate relation between the equilibrium ensemble approach
to disordered systems and conventional low concentration expansions for
such systems. It should be noted that we were as yet not forced to specify
the physical meaning of (low) concentration in the above considerations.
Clearly, this will depend on which meaning we decide to associate with
the $k_a=1$ states of the disordered system. Think for example of a
randomly spin--diluted magnet. If we choose $k_a=1$ to designate that
site $a$ is occupied by a spin, then $\r$ has the meaning of a (spin)
density, and (\ref{fexpans}) gives indeed a conventional low concentration
expansion. We might, of course, also have chosen $k_a=1$ to designate an
empty site (occupation by a vacancy), thereby generating an expansion about
the pure, homogeneous reference system. Clearly these different choices are
giving entirely different meanings to the expressions appearing in
(\ref{solvex}), (\ref{fexpans}) and (\ref{lowconc}). The quantity
$x_\emptyset = \ln Z(\emptyset)$, for instance, gives the free energy
of the empty lattice, if $k_a=1$  denotes occupancy by a spin, whereas
it gives the free energy  of the fully occupied homogeneous system, if
$k_a=1$ denotes that site $a$ is empty.

Second, we can use the formal results obtained above to compute at least
the simplest coupling constants of the exact disorder potential explicitly.
This turns out to provide useful information for the investigation of the
approximative systems introduced in the previous subsection.

Third, for models with bond-- or site--dilution, we will be able to show in
Sec.~3.2 below, that a large nontrivial
class of coupling constants of the exact disorder potential  $\phi$ will
vanish in the absence of symmetry-breaking fields. This piece of
information, too, can be useful as a guide for finding efficient sequences
of approximations within the scheme presented in Sec.~2.4.

\subsection{Models with Bond-- and Site--Dilution}

Models with bond-- and site--dilution exhibit a simplifying feature which
allows to carry the formal developement of the equilibrium ensemble
approach one step further, namely the notion of independent
non--interacting clusters in terms of which every disorder configuration
can be characterized.

If we adopt the convention that $k_a=1$ denotes an occupied site (bond), so
that the equilibrium ensemble approach generates a conventional low
concentration expansion as explained in the previous subsection, we can
establish the following results:

\begin{description}
\item[(i)] $x_\w=1$, hence $\l_\w =0$, if $\w$ denotes a {\it
disconnected}\/ set of occupied sites (bonds).
\item[(ii)] In the absence of symmetry breaking fields, $x_{\w \cup
\{a\}}=1$, hence $\l_{\w\cup\{a\}}=0$, if $\w$  denotes a connected cluster
containing at least two occupied sites (one occupied bond) to which a single
site (bond) $a$ is added to form a {\it dangling}\/ connection with $\w$.
\item[(iii)] In the absence of symmetry breaking fields, $x_{\w_1 \cdot
\w_2}=1$, hence $\l_{\w_1\cdot\w_2}=0$, if $\w_1\cdot\w_2$
denotes a connected cluster of sites (bonds) which consists of two separate
sub--clusters $\w_1$ and $\w_2$ that are conneted only via a single
common site.
\end{description}

Note that (ii) can in fact be read as a special case of (iii). All three
results hold for site or bond diluted models with nearest neighbour
interactions on arbitrary graphs. They follow directly from the
definition (\ref{defx}) of the $x_\w$. We establish them separately for
systems with site--dilution and with  bond--dilution.

\subsubsection{Models with site--dilution}

Identifying the configurations $\k$ of the disordered system with the
sets of occupied sites, we have
\be
Z(\emptyset) = x_\emptyset = 1\  ,
\ee
and
\be
Z(\k) = \prod_{\emptyset\ne\w\subseteq\k} x_\w
\ee
for $\k\ne\emptyset$. The proof of (i) is by induction. First, using
$Z(\{a\}) = x_{\{a\}} = Z_1$ for configurations in which only a single site
$a$ is occupied, and $Z(\{a\}\cup\{b\}) = Z(\{a\}) Z(\{b\})  =Z_1^2$, if
$a$ and $b$ are not nearest neighbours, one immedieately obtains
$x_{\{a,b\}}=1$ for the simplest disconnected set $\{a,b\}$ containing only
two sites. Supposing now that (i) has been proven for {\it all\/}
disconnected sets $\w$ containing up to $n$  sites, and assuming that $\k$
denotes a disconnected site containig $n+1$ sites, one has
\be
Z(\k) = \prod_{\k_\a\subseteq\k} Z(\k_\a)\ ,
\ee
where the product is over all independent clusters (maximally connected
subsets) of $\k$. By supposition
\be
Z(\k)= \prod_{\k_\a\subseteq\k} Z(\k_\a)  = \prod_{\k_\a\subseteq\k} \Big
( \prod_{\w_\a\subseteq\k_\a} \w_\a \Big )\ x_\k
\ee
entailing $x_\k = 1$ also for the disconnected set $\k$ containing $n+1$
sites, which proves (i) for the case of site--dilution.

Next, to prove (ii) we consider configurations of the form $\k = \w \cup
\{a\}$, where the site $a$ forms a {\it dangling connection}\/ with $\w$.
Denote by $E(\s_a,\s_b)$ the interaction energy between the degree of
freedom sitting on the dangling site $a$ and $\s_b$, the degree of freedom
in $\w$ to which it is connected, in the absence of symmetry breaking
fields. This interaction energy is assumed to be invariant under
an internal symmetry group of the model in the sense that
\be
\hat Z = \sum_{\s_a} \exp[-\b E(\s_a,\s_b)]
\ee
is  {\it independent}\/ of $\s_b$ so that the partition function $Z(\k)=
Z(\w\cup\{a\})$ in the absence of symmetry breaking fields is given by
\be
Z(\k)=Z(\w\cup\{a\})=\hat Z\ Z(\w)\ .
\label{factorize}
\ee
The proof of (ii) is likewise by induction. First, we have $Z(\{a\}) =
Z_1 = x_{\{a\}}$ and, by the factorizing property (\ref{factorize}),
$Z(\{a,b\})=Z_2 = \hat Z\ Z_1 = x_{\{a\}}^2\ x_{\{a,b\}}$ for a
configuration containing only two neighbouring sites $a$ and $b$. This
gives $x_{\{a,b\}} = \hat Z/Z_1$. Adding another dangling connection in
$\k=\{a,b,c\}$ gives $Z(\{a,b,c\})=Z_3={\hat Z}^2\ Z_1 = x_{\{a\}}^3\
x_{\{a,b\}}^2\ x_{\{a,b,c\}}$, so $x_{\{a,b,c\}}=1$, which verifies (ii)
for the smallest conceivable set of the form  $\w\cup\{a\}$ with $\w$
containing two sites. Suppose now that (ii) has been established for all
sets of the form $\w\cup\{a\}$ with a dangling connection from $\w$ to
a single site $a$ and with $\w$ containing up to $n$ sites, and let
$\k$ be of the form $\k=\tilde\w\cup\{a\}$ with $\tilde\w$ containing $n
+1$ sites. Then
\bea
Z(\k)  &=& Z(\tilde\w\cup\{a\})=\hat Z\ Z(\tilde\w) \nonumber\\
       &=& \Big( \prod_{\emptyset\ne\w\subseteq\tilde\w} x_\w \Big)
    \Big(\prod_{\emptyset\ne\w'\subset\tilde\w} x_{\w'\cup\{a\}} \Big)\
     x_{\{a\}}\ x_{\tilde\w\cup\{a\}}\ ,
\eea
so
\be
\hat Z =\Big(\prod_{\emptyset\ne\w'\subset\tilde\w} x_{\w'\cup\{a\}}\Big)\
x_{\{a\}}\ x_{\tilde\w\cup\{a\}}\ .
\ee
By supposition all $x_{\w'\cup\{a\}}$ in the above product except
$x_{\{a,b\}}$ are unity, where $b$ denotes the unique site in $\tilde\w$
connected to $a$ via the dangling connection. Since $x_{\{a,b\}}\ x_{\{a\}}
= \hat Z$, we can conclude that $x_{\tilde\w\cup\{a\}}=1$ as well, which
completes the proof of (ii) for the site--diluted system.

To verify (iii), note that the result $x_{\{a,b,c\}}=1$ for three
neighbouring sites $a,b,c$ obtained above establishes (iii) for the
smallest conceivable set of the form $\w_1\cdot\w_2$, with $\w_1=\{a,b\}$
and $\w_2 = \{b,c\}$. The result follows by induction after noting that
in the absence of symmetry breaking fields $Z(\w_1\cdot\w_2) = Z_1^{-1}
Z(\w_1) Z(\w_2)$, and by considering the consequences of this identity
in the $x_\w$--representation.

\subsubsection{Models with bond--dilution}

In the bond--diluted case, we identify $\k$ with the set  of occupied
bonds. Thus, if $\k=\emptyset$, the system consists of a collection of $N$
non--interacting degrees of freedom, and hence
\be
Z(\emptyset) = Z_1^N\ ,
\ee
where, as before, $Z_1$ denotes the parition function of an isolated degree
of freedom. If $\k\ne\emptyset$, there will be a set of {\it interacting\/}
degrees of freedom --- those sitting in clusters formed by bonds occupied
in $\k$ --- but in general also a collection of non--interacting degrees of
freedom. Denoting by $Z_\k$ the partition function of the degrees of
freedom {\it interacting\/} through the set of bonds occupied in $\k$, we
can write
\be
Z(\k) = Z_1^N\ \frac{Z_\k}{Z_1^{n(\k)}}\equiv Z_1^N\ \tilde Z(\k)\ ,
\ee
where $n(\k)$ is the number of vertices linked by bonds occupied in $\k$.
Returning to the $x_\w$ representation (\ref{defx}), we identify
\be
x_\emptyset = Z_1^N
\ee
and
\be
\tilde Z(\k) = \frac{Z_\k}{Z_1^{n(\k)}}= \prod_{\emptyset\ne\w\subseteq\k}
x_\w\ .
\label{defxtilde}
\ee
The $\tilde Z(\k)$ have properties analogous to the $Z(\k)$ in the
site--diluted problem. In particular, if $\k=\k_1\cup\k_2$ and $\k_1$ and
$\k_2$ are mutually disconnected sets of occupied bonds, then $\tilde
Z(\k) = \tilde   Z(\k_1) \tilde Z(\k_2)$. This factorizing property
again entails $x_\w=1$, if $\w$ denotes a disconnected set of bonds.
The proof is by induction in complete analogy to the site diluted case,
and it will not be repeated here.

Next, to establish (ii) for the bond diluted case, we note that
\be
\tilde  Z(\w\cup\{a\}) = \frac{Z_{\w\cup\{a\}}}{Z_1^{n(\w)+1}} =
\frac{\hat Z}{Z_1} \frac{Z_{\w}}{Z_1^{n(\w)}} = \frac{\hat Z}{Z_1} \tilde
Z(\w)
\ee
in the absence of symmetry breaking fields, if $a$ denotes a dangling
connection added to the set $\w$; here $\hat Z$ has the same meaning
as in the site diluted case. Again, by inductive reasoning in complete
analogy with the site diluted problem, one verifies (ii), i.e.,
$x_{\w\cup\{a\}}=1$ if $\w$ contains at least one occupied bond.

Finally, to prove (iii), we note that
\be
\tilde Z(\w_1\cdot\w_2)=\frac{Z_{\w_1\cdot\w_2}}{Z_1^{n(\w_1)+n(\w_2)-1}}
=\frac{Z_{\w_1}}{Z_1^{n(\w_1)}} \frac{Z_{\w_2}}{Z_1^{n(\w_2)}}
=\tilde  Z(\w_1) \tilde Z(\w_2)\ ,
\ee
where the second equality requires the absence of symmetry breaking fields.
This factorizing property of the $\tilde Z(\w_1\cdot\w_2)$ entails (iii) on
the level of the $x_\w$--representation.

\subsubsection{One dimensional models with bond-- or site--dilution}

One dimensional systems with bond or site dilution are particularly
simple because in 1--d one has a complete overview over the family of
connected clusters which consists just of isolated (finite) chains.
This allows to give an explicit representaion of the $x_\w$ defined
through (\ref{defx}) in terms of chain free energies, which we briefly
record here.

For the site diluted problem, denoting by $Z_n$ the partition function
of an isolated open chain of $n$ occupied sites, and by $x_n$ the
corresponding $x$ variable, we have
\be
Z_n = \prod_{k=1}^n x_k^{n+1-k}\ ,
\ee
which gives $x_1=Z_1$, and (with $Z_0=1$)
\be
x_n= Z_n Z_{n-2}/Z_{n-1}^2\ ,\qquad n\ge 2\ .
\label{xnsite}
\ee
Inserting this into the expression (\ref{fexpans}) for the free energy of
the equivalent equilibrium ensemble and normalizing with respect to
sample size  $N=|\L|$, we get
\bea
-\b f^\phi &=& \lim_{N\to\infty} N^{-1}\ln Z^\phi\nonumber\\
&=& -\r\ln\r -(1-\r)\ln(1-\r) + (1-\r)^2 \sum_{n=1}^\infty \r^n \ln Z_n\ .
\label{fsitechain}
\eea
Up to the contribution coming from the entropy of mixing, this reproduces
the exact solution as obtained, e.g., by Wortis \cite{wor}. In the absence
of
symmetry breaking fields we have $\ln x_n  = 0$ for $n\ge 3$, so in this
case
\be
-\b f^\phi = -\r\ln\r -(1-\r)\ln(1-\r) + \r\ln Z_1 + \r^2 \ln(Z_2/Z_1^2)\ .
\ee
In the bond diluted problem, isolateds chain containing $n$ successive
bonds constitute chains of $n+1$ interacting degrees of freedom. If we
formulate (\ref{defxtilde}) for such chains, it reads
\be
\frac{Z_{n+1}}{Z_1^{n+1}} = \prod_{k=1}^n x_k^{n+1-k}\ ,
\ee
where the $x_k$ now refer to chains containing $k$ {\it bonds\/}. This
gives $x_1 = Z_2/Z_1^2$, and
\be
x_n = Z_{n+1}Z_{n-1}/Z_n^2\ ,\qquad n\ge 1.
\label{xnbond}
\ee
Comparing with the corresponding expression (\ref{xnsite}) for the
site--diluted case, we note that $x_n^{{\rm bond}}=x_{n+1}^{{\rm site}}$
for $n\ge 1$. For the free energy per site in the bond--diluted system
we thus get (recalling $x_\emptyset  =Z_1^N$)
\be
-\b f^\phi = -\r\ln\r -(1-\r)\ln(1-\r)
+ (1-\r)^2 \sum_{n=1}^\infty \r^{n-1} \ln Z_n\ ,
\label{fbondchain}
\ee
thereby reproducing again the exact solution\cite{wor}. In the absence of
symmmetry breaking fields, $\ln x_n = 0$ for $\n\ge 2$, and the expression
simplifies to
\be
-\b f^\phi = -\r\ln\r -(1-\r)\ln(1-\r) + \ln Z_1 + \r \ln(Z_2/Z_1^2)\ .
\ee
Eqs. (\ref{fsitechain}) and (\ref{fbondchain}) can be evaluated for all
models for which the finite chain thermodynamics is known, e.g., from
transfer--matrix techniques.

Note that the result (ii) concerning the vanishing of coupling constants
of the disorder potential of the form $\l_{\w\cup\{a\}}$ in the absence
of symmetry breaking fields implies that in this limit very simple
approximations in the moment--matching approximation scheme described in
Sec.~2.2 already provide {\it exact\/} solutions for bond--  or
site--diluted chains. For the bond--diluted case, it is the simplest
conceivable annealed approximation enforcing only the average occupancy
$\ld k_a\rd =\r$ of the bonds.  In the case of site--dilution, one has
to fix the average site--occupancy at $\r$ and the expectations
$\ld k_a k _b\rd$ for neighbouring sites at $\r^2$ in order to reproduce
the thermodynamics of the quenched system exactly.

On the contrary, if there {\it is\/} a nonvanishing symmetry breaking
field, then Eqs.~(\ref{xnsite}) and (\ref{xnbond}) imply that {\it no
finite approximation\/} in the moment--matching scheme described in
Sec.~2.2 is exact. Nevertheless, the simplest approximations enforcing
only $\ld k_a\rd=\r$ in the bond--diluted case, and both, the site
occupancy
$\ld k_a\rd=\r$ and the nearest neighbour expectation $\ld k_a k_{a+1}\rd
=\r^2$ in the site--diluted case already give rather precise results,
as will be demonstrated for the Ising model in II.

\section{Thermally Correlated Frozen-In Disorder}

Having treated uncorrelatded bimodal bond-- and site--disorder
in some detail, it is perhaps appropriate to recall that the equlibrium
ensemble approach to disordered systems is not restricted to this case.
Except in the previous section we have, in fact, never made use of the
specific simplifications pertaining to uncorrelated disorder. The moment
matching schemes explaind in Secs. 2.4--2.6. can naturally also be put
to work in the case of {\em correlated\/} disorder. The only nontrivial
piece of information needed is contained in the system of moments
(\ref{moments}) or (\ref{cmoments})  for the bimodal and non-bimodal
cases, respectively. In real physical systems, which {\em do\/} as a rule
exhibit some degree of correlations between the configurational degrees of
freedom, a subset of the full set of moments, obtained \eg\ through
scattering experiments, is often the only piece of information that is
actually available about the disordered system in question. So, in
situations where relatively simple approximations in a moment matching
scheme can be expected to produce reliable results, the equilibrium
ensemble approach appears to be an ideally suited tool to use.

In the present section we will treat {\em correlated}\/ bimodal bond--
and site--disorder in some detail. We will restrict our attention to the
case where the correlations are produced as a result of some previous
annealing proccess (in the narrow metallurgical sense). That is, we suppose
that the quenched disorder distribution is an equilibrium distribution
produced at values of the external parameters (temperature, fields,
chemical potential, etc.) characteristic of the preparation process, but
in general different from the values at which experiments on the
system are actually being perfomed.

Let us assume that the system is prepared at (inverse) temperature $\bar\b$
and that it is described by the total Hamiltonian
\be
\bar\cH_{\rm tot}(\s,\k) = \bar\cH(\s|\k) + V(\k)
\ee
where $\bar\cH(\s|\k)$ is the Hamiltonian of the disordered system
describing the energy of the $\s$--degrees of freedom at fixed disorder
configuration $\k$, albeit at parameter settings characteristic of the
preparation process, which we indicate by the overbar. The potential
$V(\k)$ accounts for the fact that there may be an additional interaction
energy between the configurational degrees of freedom which produces
correlations between the $k_a$ over and above that mediated by $\s$. Think,
for example, of a dilute magnetic alloy. The distribution of particles will
usully be governed, both, by chemical interactions --- accounted for by
$V(\k)$ ---  and by $\k$--dependent interactions between magnetic degrees
of freedom described by $\bar\cH(\s|\k)$.  In typical cases, the former
may, in fact, be expected to dominate.

We assume that during the preparation process the temperature is high
enough to allow the system to come to complete equilibrium. The distribution
of the configurational degrees of freedom is then given by
\be
\bar q(\k) = \frac{\bar Z(\k)}{\bar Z} \exp[-\bar\b V(\k)]\ ,
\label{barq}
\ee
where $\bar Z(\k)=\sum_\s \exp[-\bar\b \bar\cH(\s|\k)]$. If the system is
then cooled down to lower temperatures, the impurity distribution
(\ref{barq}) may be regarded as fixed, and the disordered system
characterized by this quenched distribution will exhibit frozen
correlations between impurities. In what follows, we will show that the
formal developments presented in Sec. 3 for the uncorrelated case can
be carried almost as far in the case of thermally correlated quenched
disorder.

To this end we return to the Morita equation (\ref{solvephi}) and the
representation
\be
\b \phi(\k) = - \sum_{\emptyset \ne\w\subseteq\L} \l_\w ( k_\w(\k) - \bar f_w )
\ee
of the disorder potential $\phi(\k)$, the overbar indicating that the
moments are evaluated with (\ref{barq}).

The coupling constants of the disorder potential for this problem
of correlated disorder can now be determined along the same lines as in
Sec.~3.1. For the bimodal disorder type considered here, we can always
write
\be
V(\k) = \sum_{\w\subseteq \L} v_\w k_\w(\k)
\ee
Using the $x$ representation (\ref{defx}) for both $Z(\k)$ and for the
partition function $\bar Z(\k)$ with the parameters settings prevalent
during the preparation process, we get
\be
\l_\w = -\bar\b v_\w - \ln  (x_\w/\bar x_\w)\quad ,\quad \w\ne\emptyset\ .
\ee
Returning to the $\k=\emptyset$--version of the Morita equation, and
recalling that $\bar f_\emptyset = 1$ irrespectively of the correlations,
and that $Z(\emptyset)=x_\emptyset$ and similarly $\bar Z(\emptyset) =
\bar x_\emptyset$, the free energy $F^\phi$ is obtained in the form
\bea
-\b F^\phi & = & \ln Z^\phi = \ln \bar Z + \sum_{\w\subseteq\L} \big (
\bar\b v_\w + \ln(x_\w/\bar x_\w)\big ) \bar f_\w \nonumber \\
& = & -\ld \ln \bar q(\k)\rd_{\bar q} + \sum_{\w\subseteq\L} \ln x_\w \bar
f_w\ .
\label{barfexpans}
\eea
The reader is invited to compare this with (\ref{fexpans}). The main
additional obstacle on the way to evaluating the moment expansion
(\ref{barfexpans}) to high orders --- assuming such an evaluation were
possible for the concentration expansion(\ref{fexpans}) in the uncorrelated
case --- lies in the $\bar f_\w$, which cannot be expected to be of such a
simple form as in the case of uncorrelated disorder. Nevertheless, analytic
approximations are conceivable in the form of high--temperature expansions
of the $\bar f_\w$, if $\bar\b$ is sufficiently small. In any case, as will
be shown in II, relatively simple moment matching schemes, that is, simple
approximations in the restricted annealing approach, which do not require
knowledge of high moments, may still be feasible and useful.

It should be noted that the disorder potential $\phi(\k)$ reduces to
$V(\k)$, if parameter values are the same as those during system
preparation, so that $x_\w=\bar x_\w$. The reader may convince herself that
this just what is needed produce the correct limiting form of $Z^\phi$ in
the case where $\bar q(\k)$ is the equilibrium distribution of the
configurational degrees of freedom $\k$.

Specializing to systems  with bond-- or site--dilution, we observe that
the results concerning the vanishing of certain coupling constants of
the disorder potential $\phi$ stated and proved in Sec.~3.2 have simple
analogues in the case of thermally correlated disorder. To see this, note
that the results (i)--(iii), if formulated on the level of the $x_\w$,
do not depend on disorder statistics or parameter settings; so they hold
for the $\bar x_\w$ alike.  For thermally correlated quenched disorder,
statements to the end that $x_\w = \bar x_\w =1$ then translate into
$\l_\w = -\bar\b v_\w$. This implies a vanishing of the coupling constants
as in the uncorrelated case, if the the potential $V(\k)$ involves no
interactions beyond those coupling configurational degrees of freedom on
neighbouring sites (bonds).

Taking site--diluted chains in the absence of symmetry breaking fields
as an example, we obtain the following surprisingly simple expression
for the free energy per site as computed in the equilibrium ensemble
approach,
\be
-\b f^\phi = k_{\rm B}^{-1} \bar s_0 + \bar f_1 \ln Z_1 + \bar f_2 \ln
(Z_2/Z_1^2)\ ,
\label{barsite}
\ee
where $\bar s_0$ is the entropy of mixing per site, while $\bar f_1 =
\ld k_a\rd_{\bar q} =\r$ and $\bar f_2 =\ld k_a k_{a+1}\rd_{\bar q}$.
That is all effects of quenched correlations on the free energy of the
system make themselves felt {\em only\/} through the modified nearest
neighbour correlation. This ceases to be the case if a symmetry breaking
field is applied. Note that symmetry breaking fields during the preparation
process manifest themselves in this context --- beyond their effect in
modifying $\bar f_1$ and $\bar f_2$ --- only in modifying the entropy of
mixing.

Similarly, in the bond--diluted system with thermally correlated quenched
bond occupancy one obtains
\be
-\b f^\phi = k_{\rm B}^{-1} \bar s_0 + \ln Z_1 + \bar f_1 \ln
(Z_2/Z_1^2)\ .
\label{barbond}
\ee
So the result depends only on the average occupancy $\bar f_1=\r$ of the
bonds and is {\em independent\/} of correlations between them.

The simplicity of the zero--field solutions (\ref{barsite}) and
(\ref{barbond}) for the 1--D systems with quenched thermal correlations
is not easily detected in the exact solution \cite{kuc}, which has been
obtained by the direct averaging approach.

\section{Concluding Remarks}

We have presented an outline of the formal and systematic aspects of
Morita's equilibrium ensemble approach to systems with quenched disorder
\cite{mor}. Hitherto unnoticed relations to other, more conventional
sets of ideas, such as low concentration expansions, perturbation
expansions about pure reference systems, variational bounding of free
energies, and generalizations of the idea of grand ensembles have been
pointed out and elaborated.

The {\em canonical\/} approximation scheme within Morita's approach ---
a moment matching procedure that goes under the name of restricted or
constrained annealing --- is of a {\em non--perturbative\/} nature, which
should also be clear in view of its relation to variational methods.
Conversely, the expansion of the {\em exact\/} disorder potential was
demonstrated to generate conventional perturpation expansions. A third,
in some sense intermediate approach would be to attempt solving or
analysing models with a truncated expansion of the full disorder potential
exactly.

For models with bond-- or site--dilution, it was demonstrated that a large
non--trivial class  of coupling constants of the exact disorder potential
$\phi(\k)$ vanishes in the absence of symmetry--breaking fields. This fact
may help to explain why relatively simple approximations  within the moment
matching schemes or the variational formulation presented in Sec.~2 have
been so successful in describing aspects of critical behaviour of \eg\
disordered Ising models ---  undisputably so at least regarding the
computation of phase transition lines (\cite{tb}, \cite{tho}, \cite{geo+},
\cite{kuprl}), still under debate though (see, \eg, \cite{prlc}), regarding
the much more controversial and difficult question of critical exponents
\cite{kuprl}, \cite{II}.

In the case of continuous disorder distributions, straightforward moment
matching schemes as explained in Sec.~2 might not always prove to be
the most efficient way of formulating approximative solutions. Instead
of fixing moments up to a certain order directly, one might, for instance
think of reproducing expectations only of certain {\em combinations\/}
of moments of the quenched disorder distribution in the equilibrium
ensemble. The choice of particular combinations might be guided by attempts
to exploit powerful analytic structures, such as provided by orthogonal
function systems, or by physical insight. In bond--disordered Ising
models,
for instance, the representation of the exact solution in terms of weighted
sums over configurations of van der Waerden polygons suggests that the
decisive quantities to be reproduced in an equilibrium ensemble approach
would be gauge--invariant loop-correlations of the natural
high--temperature variables $t_{ij}=\tanh(\b J_{ij})$, expressed here in
terms of the conventional notation for the exchange couplings. That is,
the disorder potential $\phi(\k)\equiv \phi(\{t_{ij}\})$ would be designed
to fix correlations of the $t_{ij}$ aroud closed loops $\cC$, \ie,
correlations of the form $\ld \prod_{(i,j)\subset\cC} t_{ij}\rd$  at
their corresponding quenched values, each such correlation involving
moments of the corresponding quenched $\{J_{ij}\}$--distribution of
arbitrarily high order. This idea has been exploited by George \ea\/
\cite{geo+} to locate phase boundaries in disordered Ising models, and
it is also being dicussed in a recent preprint of Paladin \ea\/
\cite{sepab}.

Finally, it is perhaps worth pointing out once more that the equlibrium
ensemble approach appears to be well suited to study problems with
{\em correlated\/} disorder, as is borne out by the general theory of
Sec.~2, and by the more specific considerations of Sec.~4. Within the
framework of (low order) moment matching schemes at least, correlated
disorder appears to be hardly more complicated than uncorrelated disorder.
Regarding more conventional approaches, this can --- at best --- be claimed
for correlated disorder of the Gaussian type.

{\bf Acknowlegements:} The author is indebted to A. Huber for having
acquainted him with Morita's ideas. Illuminating discussions with  B.
Derrida, H. Horner, R. Stinchcombe, J. Vannimenus, and F. Wegner
are gratefully acknowledged. It is a pleasure to thank the members of
the Instituut voor Theoretische Fysica at K.U. Leuven for the hospitality
extended to him, while parts of this paper were being written, and D.
Boll\'e and J. van Mourik for helpful comments.

\end{document}